\documentclass[amsmath,twocolumn,amssymb,prb,aps,superscriptaddress,letterpaper,showpacs]{revtex4}

\usepackage{graphicx}
\usepackage{dcolumn}
\usepackage{bm}
\usepackage{amsmath}

\begin{document}

\title{Interplay of the Aharonov-Bohm effect and Klein tunneling in graphene}

\author{J{\"o}rg Schelter}
\affiliation{Department of Theoretical Physics and Astrophysics,
University of W{\"u}rzburg, D-97074 W{\"u}rzburg, Germany}

\author{Dan Bohr}
\affiliation{Department of Physics and Astronomy,
University of Basel, CH-4056 Basel, Switzerland}

\author{Bj{\"o}rn Trauzettel}
\affiliation{Department of Theoretical Physics and Astrophysics,
University of W{\"u}rzburg, D-97074 W{\"u}rzburg, Germany}

\date{\today}

\begin{abstract}
We numerically investigate the effect of Klein tunneling on the Aharonov-Bohm oscillations in graphene rings using a tight-binding model with nearest-neighbor couplings. In order to introduce Klein tunneling into the system, we apply an electrostatic potential to one of the arms of the ring, such that this arm together with the two adjacent leads form either a $nn'n$- or $npn$-junction ($n,n'$: conduction band transport, $p$: valence band transport). The former case corresponds to normal tunneling and the latter case to Klein tunneling. We find that the transmission properties strongly depend on the smoothness of the $pn$-interfaces. In particular, for sharp junctions the amplitude profile is symmetric around the charge neutrality point in the gated arm, whereas for smooth junctions the Aharonov-Bohm oscillations are strongly suppressed in the Klein tunneling as compared to the normal tunneling regime.
\end{abstract}


\maketitle

\section{Introduction}\label{sec:intro}

It is by now common knowledge that graphene has peculiar transport properties making the material an interesting candidate for future applications (for recent reviews on graphene see Refs.~\onlinecite{Geim2007,Been2008,Cast2009}). Ballistic transport in graphene has been coined pseudo-diffusive at the Dirac point \cite{Kats2006,Twor2006,Prad2007} because it is carried by evanescent modes. This gives rise to transport properties that resemble diffusive transport in other materials -- a prediction that has been experimentally confirmed in shot noise measurements.~\cite{Dica2008,Dann2008} Phase-coherent transport in disordered graphene is not less interesting than ballistic transport in clean graphene. The reason is that the honeycomb lattice of graphene in combination with different type of scattering mechanisms yields rather rich localization physics.~\cite{Mcca2006} Depending on the magnitude of the so-called intervalley scattering, one can either see weak antilocalization or weak localization.~\cite{Morp2006} A recent experiment on quantum interference in graphene has confirmed this prediction by measuring the same sample at different carrier densities and temperatures which allows to see the transition from localization to antilocalization.~\cite{Tikh2009}

\begin{figure}[t]
\centering
\includegraphics[width=0.95\columnwidth]{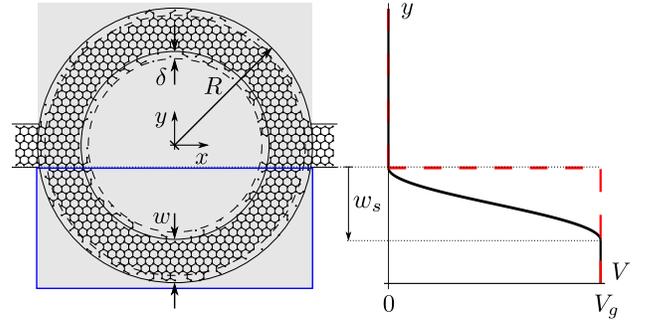}
\caption{(Color online) Schematic of the graphene ring structure (left) and the $y$-dependence of the on-site gate potential $V$ (right) that is applied to the lattice sites on the lower arm of the ring. $V$ is non-zero only within the blue box and exhibits either a smooth (black solid line) or sharp (red dashed line) profile along $y$-direction while being constant along $x$-direction. The width of the arms of the ring is chosen equal to the width $w$ of the leads. The shaded area indicates the region of non-vanishing homogeneous magnetic field pointing out of plane. Different edge disorder configurations are realized by randomly removing sites within the two regions of width $\delta$ between dashed and solid circles.}
\label{fig:setup}
\end{figure}

Therefore, the combination of ballistic transport at (or close to) the Dirac point with quantum interference effects suggests itself to contain interesting physics. This is our motivation to study how the Aharonov-Bohm effect \cite{Ahar1959} in graphene rings is affected by tuning one of the arms of the ring with an external gate through the Dirac point, see Fig.~\ref{fig:setup} for a schematic. We will show below that such a setup allows for a clear graphene-specific signature in Aharonov-Bohm measurements which seems to be readily observable. Its physical origin is the quantum interference of normal tunneling as well as Klein tunneling trajectories through the two arms of the ring.

Previous theoretical work on graphene ring structures includes the investigation of the electronic properties of closed single-layer~\cite{Rech2007} and bi-layer~\cite{Zare2009} rings pierced by a magnetic flux as well as transport studies of the Aharonov-Bohm effect in clean and disordered graphene away from the Dirac point.~\cite{Ryce2009,Wurm2009} It should be mentioned that none of the graphene-specific predictions of Refs.~\onlinecite{Rech2007,Zare2009,Ryce2009,Wurm2009} have been observed in subsequent Aharonov-Bohm oscillation measurements.~\cite{Russ2008,Huef2009} Recently, Katsnelson has studied the Aharonov-Bohm effect in undoped graphene (at the Dirac point) and made graphene-specific predictions that are complementary to ours and might be observable in future experiments.~\cite{Kats2010}

The article is organized as follows. In Sec.~\ref{sec:model}, we introduce the tight-binding model that we use for the transport analysis. This section includes a brief description of the recursive Green's function formalism with an emphasis on the peculiarities due to graphene's honeycomb lattice. Subsequently, in Sec.~\ref{sec:results}, we discuss our results in the different transport regions which show the interplay of the Aharonov-Bohm effect and Klein tunneling in phase-coherent graphene nanostructures. Finally, we conclude in Sec.~\ref{sec:con}. Some technically details are discussed in the appendices.

\section{Model}\label{sec:model}

Our calculation starts with the usual tight-binding Hamiltonian for graphene
\begin{equation}
	\mathcal{H} = \sum_i{ V_i \left| i \right\rangle \left\langle i \right| } + \sum_{ \left\langle i, j \right\rangle }{ \tau_{ij} \left| i \right\rangle \left\langle j \right| }
\end{equation}
where the second sum runs over nearest-neighbors and $V_i = V(\mathbf{r}_i)$ is an on-site potential that may depend on position. The graphene hopping integral $\tau_0 \sim 3 \, \mathrm{eV}$ picks up a Peierls phase in the presence of a magnetic field yielding for the nearest-neighbor coupling element the expression
\begin{equation}
	\tau_{ij} = -\tau_0 \, \exp{ \left( \frac{ 2 \pi \mathrm{i} }{ \Phi_0 } \int_{ \mathbf{r}_i }^{ \mathbf{r}_j }{ \mathbf{A}(\mathbf{r}) \, \mathrm{d}\mathbf{r} } \right) }
\end{equation}
where the line integral is taken along the straight path between sites $i$ and $j$. $\Phi_0 = h / e$ is the magnetic flux quantum.

The system under consideration is a ring-shaped structure cut out of a graphene sheet, which is attached to two crystalline leads also modeled using the graphene lattice structure (see Fig.~\ref{fig:setup}). The structure is subject to a homogeneous magnetic field $\mathbf{B}(\mathbf{r}) = (0, 0, B)$ resulting from a vector potential $\mathbf{A}(\mathbf{r}) = (-B \, y, 0, 0)$ as well as a gate electrode potential $V_g$ located on top of the lower arm of the ring. The smoothness of the potential interface is controlled via the smoothing width $w_s$ measured from the lower edges of the leads:
\begin{align*}
	V = 0 \quad &\mathrm{for} \quad y \geq -w / 2,\\
	V = V_g	 \quad &\mathrm{for} \quad y \leq -w / 2 - w_s,\\
	0 < V < V_g \quad &\mathrm{otherwise},
\end{align*}
taking the origin of coordinates at the center of the ring. In our simulations, we used a cosine-shaped smoothing profile and chose values $w_s = 0 \ldots R - 3 w / 2$.

\begin{figure}[t]
\centering
\includegraphics[width=0.95\columnwidth]{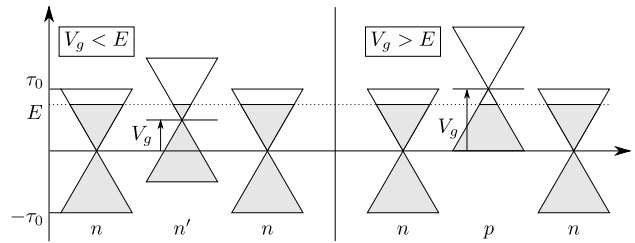}
\caption{Schematic of the influence of the potential profile introduced by $V_g$ on the spectrum of the lower arm of the ring. The left hand side shows the normal tunneling case ($nn'n$-junction) and the right hand side the Klein tunneling case ($npn$-junction). The location of the van Hove singularities at $\pm \tau_0$ is also schematically depicted.}
\label{fig:junctionDirac}
\end{figure}

For a Fermi energy $E > 0$, together with the adjacent leads this lower arm forms either a $nn'n$- or $npn$-junction for $V_g < E$ and $V_g > E$, respectively (see Fig.~\ref{fig:junctionDirac} for a schematic). Note that the setup exhibits a flat potential profile for trajectories along the upper ring arm, i.e. a $nnn$-junction, since there is no gate potential applied. This enables a rather large transmission through the ring even when the lower ring arm is tuned towards the Dirac point, since transport through the upper arm always takes places at an energy distance $E$ away from the charge neutrality point.

We derive the transmission function through the ring from the scattering $S$-matrix using the Landauer-B{\"u}ttiker formalism for elastic transport at zero temperature assuming complete phase-coherence. The semi-infinite left and right leads are described through their respective real-space surface Green's functions (SGF) $g_L$ and $g_R$. The coupling of the leads via the Hamiltonian $\mathcal{H}$ describing the ring structure is expressed through Dyson's equation. In order to obtain the effective coupling between the leads, we apply a recursive Green's function (RGF) technique. The $S$-matrix is then obtained from the system's Green's function using the Fisher-Lee relation.~\cite{Fish1981}

\subsection{Calculation of the lead's surface Green's function}

For the calculation of the lead's surface Green's functions, we essentially follow Ref.~\onlinecite{Sanv1999}. An isolated infinite lead is described as quasi-one-dimensional periodic arrangement of identical unit cells. Each unit cell is described by an intra-cell matrix $H_0$ whose dimension equals the number $M$ of atomic sites within the unit cell. Neighboring unit cells are coupled via an inter-cell matrix $H_1$, such that for each cell a Schr{\"o}dinger equation of the form
\begin{equation}
	H_1^\dagger \, \psi_{ z - 1 } + H_0 \, \psi_z + H_1 \, \psi_{ z + 1 } = E \, \psi_z, \qquad z \in \mathbb{Z}
\end{equation}
holds, where $\psi_z$ is a $M$-dimensional vector describing the atomic sites within a particular unit cell. Since we consider the zero temperature regime, transport takes place at the Fermi energy $E$, measured relative to the charge neutrality point in the leads. The approach outlined in Ref.\ \onlinecite{Sanv1999} then requires the inversion of the inter-cell matrix $H_1$. However this matrix is singular in the case of a graphene lattice. To overcome this problem, we will use an effective description for the leads as outlined in App.~\ref{sec:app_effLeads}. This description is based on the fact, that in many cases not all of the atomic sites contained in one cell couple to the adjacent cell. One can then conveniently divide the unit cells into even- and odd-indexed subcells and eliminate the latter from the calculation by writing down the effective coupling between even-indexed subcells. This yields an expression of the form
\begin{equation}
\label{for:effSG}
	\tilde{H}_1^\dagger \, \tilde{\psi}_{ z - 1 } + \tilde{H}_0 \, \tilde{\psi}_z + \tilde{H}_1 \, \tilde{\psi}_{ z + 1 } = E \tilde{\psi}_z
\end{equation}
with invertible coupling matrices $\tilde{H}_1$. As will be shown later, this effective description also has the additional benefit of increased performance of the recursive Green's function scheme. Note that since the approach does not depend on details of the matrices $H_0$ and $H_1$ it may also be applied to other lattice structures that exhibit non-invertible coupling matrices or that enable a potential performance gain.

The isolated leads' SGFs $g_L$ and $g_R$ are thus obtained according to expressions (2.16) and (2.17) of Ref.\ \onlinecite{Sanv1999}, respectively, using the effective description (\ref{for:effSG}) for the leads.

\subsection{Connecting leads and conductor}

In order to obtain the $S$-matrix, we need to compute the Green's function
\begin{equation}
\label{for:fullGF}
G = \left( \begin{array}{cc}
  G_{11}		&	G_{12}\\
  G_{21}		&	G_{22}
\end{array} \right) =
\left( \begin{array}{cc}
  \tilde{g}_L		&	\tilde{g}_{LR}\\
  \tilde{g}_{RL}		&	\tilde{g}_R
\end{array} \right)
\end{equation}
describing the effective coupling between the surface sites of the two leads. To this end, we apply a variant of the commonly-used RGF scheme.~\cite{Lee1981,McKi1985,Todo1994,Lake1997} As described in App.~\ref{sec:app_interface}, in an effective description, the leads have to couple to the conductor through an additional contact slice, which has to be taken into account in contrast to conventional RGF algorithms.

Similar to the leads, the sample itself is also divided into slices. The width $\Delta x$ of the slices is chosen as small as possible, since the algorithm scales only linear with the length of the sample (along $x$-direction) but up to third power with the width (along $y$-direction). We therefore choose $\Delta x = a_x / 2$ for zigzag leads and $\Delta x = a_x / 4$ for armchair leads, where $a_x$ is the lattice constant of the honeycomb lattice along the $x$-direction, which is $a_x = a_0 \sqrt{3}$ in the zigzag case and $a_x = 3 a_0$ in the armchair case, $a_0 = 0.142 \, \mathrm{nm}$ being the nearest-neighbor distance in graphene.

The applied RGF procedure then has the following structure:
\begin{enumerate}
\item Connect the contact slices to the initially isolated leads.
\item Connect to slice $1$ of the conductor.
\item Set $n = 1$.
\item Connect to slice $n + 1$ of the conductor and eliminate slice $n$ from the description.
\item Increase $n$ by one and repeat the previous step until all slices are connected.
\end{enumerate}
In each step it is sufficient to update the contact slice Green's functions. The Green's function (\ref{for:fullGF}) of the fully coupled leads is updated once when all slices are connected.~\cite{foot1} The scheme of a particular step in the recursion is as follows: Interpreting the coupling to slice $n$ as perturbation to the system of coupled slices up to slice $n - 1$, the Green's function describing the coupled system
\begin{equation}
\label{for:RGF_fullGF}
g(n) \equiv \left(\begin{array}{ccc}
g_C(n)		&	g_{CX}(n)	&	g_{CD}(n)\\
g_{XC}(n)	&	g_X(n)		&	g_{XD}(n)\\
g_{DC}(n)	&	g_{DX}(n)	&	g_D(n)
\end{array}\right)
\end{equation}
is obtained from the unperturbed Green's function
\begin{equation}
\label{for:isolatedGF}
g_0(n) \equiv
\left(\begin{array}{ccc}
g_C(n-1)	&	g_{CD}(n-1)	&	0\\
g_{DC}(n-1)	&	g_D(n-1)		&	0\\
0			&	0			&	(E - H^{(n)})^{-1}
\end{array}\right)
\end{equation}
and the coupling matrix (the perturbation)
\begin{equation}
\mathcal{U} \equiv
\left(\begin{array}{ccc}
0						&	0						&	T_{0, n}\\
0						&	0						&	T_{n-1,n}\\
\left(T_{0, n}\right)^\dagger	&	\left(T_{n-1,n}\right)^\dagger	&	0
\end{array}\right)
\end{equation}
via Dyson's equation
\begin{equation}
g(n) = g_0(n) + g_0(n) \, \mathcal{U} \, g(n).
\end{equation}
Here, the index $C$ refers to the contact slices, whereas the index $D$ refers to the last connected slice of the sample. $H^{(n)}$ is the layer-local Hamiltonian of slice $n$ of the sample, and the matrices $T_{m, n}$ describe the coupling of the contact slices and slice $n - 1$ to slice $n$ of the sample. For a simple two-terminal setup as we consider here, $T_{0, n}$ is non-zero only for $n \in \{ 1, N \}$, $N$ being the total number of slices of the sample.

Note that the matrix elements containing an index $X$ in (\ref{for:RGF_fullGF}) do not have to be calculated since they do not appear in the next step of the recursion. Further note that we do not need to add an inconvenient infinitesimal imaginary part to the energy in expression (\ref{for:isolatedGF}) since it would anyway be absorbed by finite imaginary terms introduced by the fact that we deal with an open quantum system.

\subsection{Calculation of the scattering matrix}

The linear conductance of the system is obtained using the Landauer formula
\begin{equation}
\mathcal{G} = \mathcal{G}_0 \ \mathrm{Tr}(t^\dagger t)
\end{equation}
where $\mathcal{G}_0 \equiv 2 e^2 / h$. The factor $2$ accounts for spin degeneracy and $t$ is the transmission matrix element of the scattering matrix
\begin{equation}
S = \left( \begin{array}{cc}
  S_{11}	&	S_{12}\\
  S_{21}	&	S_{22}
\end{array} \right) = \left( \begin{array}{cc}
  r	&	t'\\
  t	&	r'
\end{array} \right)
\end{equation}
which itself is a matrix whose elements are the transmission amplitudes for scattering between the different transverse modes in the two leads. The $S$-matrix can be written in terms of the Green's function $G_{ij}$ in Eq.~(\ref{for:fullGF}) by means of the Fisher-Lee relation \cite{Fish1981}
\begin{equation}
\label{for:FL}
(S_{ij})_{hl} = \tilde{\phi}_{\bar{h}}^\dagger \ \left( -\delta_{ij} \cdot \mathbf{1} + G_{ij} \mathcal{V} \right) \ \phi_l \cdot \sqrt{v_h / v_l} .
\end{equation}
In the latter equation, the transverse eigenvectors $\phi_l$ of in-moving states in lead $j$ and the duals $\tilde{\phi}_{h}^\dagger$ of the out-moving states in lead $i$ as well as the corresponding group velocities $v_{h, l}$ and the matrix $\mathcal{V}$ are defined and calculated as described in Ref.~\onlinecite{Sanv1999}.

\section{Results}
\label{sec:results}

In the following we present transmission properties for a ring with $R / a_0 = 300$ and $w / a_0 = 60$. Edge disorder is applied to the ring by randomly removing sites within a width $\delta$ from the inner and outer edges of the ring, respectively (see Fig.~\ref{fig:setup}). We choose $\delta / a_0 = 1.5$ in order to keep the edge of the ring as smooth as possible while still allowing for different edge disorder configurations. As depicted in Fig.~\ref{fig:junctionDirac}, Fermi energy $E \in \left] 0, \ldots, \tau_0 \right[$ and gate potential $V_g \in \left[ 0, \ldots, 2 \, E \right]$ are chosen such that transport always takes place in between the van Hove singularities located at $E = \pm \tau_0$ where the density of states diverges in the tight binding model of graphene.

\begin{figure}[t]
\centering
\includegraphics[width=0.95\columnwidth]{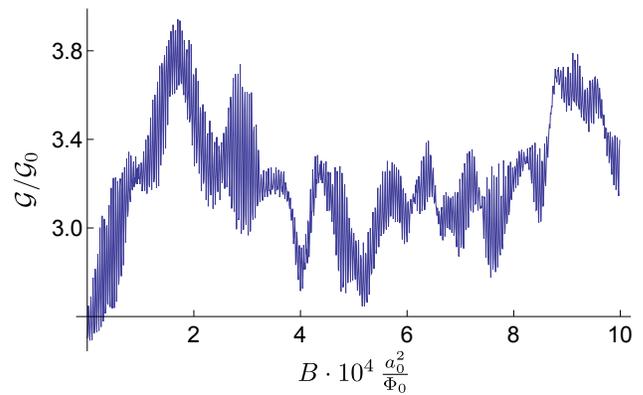}
\caption{(Color online) Magnetoconductance of a ring with $R / a_0 = 300$, $w / a_0 = 60$ at energy $E / \tau_0 = 0.5$ and zero gate voltage, showing clear Aharonov-Bohm oscillations on top of a background due to universal conductance fluctuations.}
\label{fig:magnetoconductance}
\end{figure}

\begin{figure}[t]
\centering
\includegraphics[width=0.95\columnwidth]{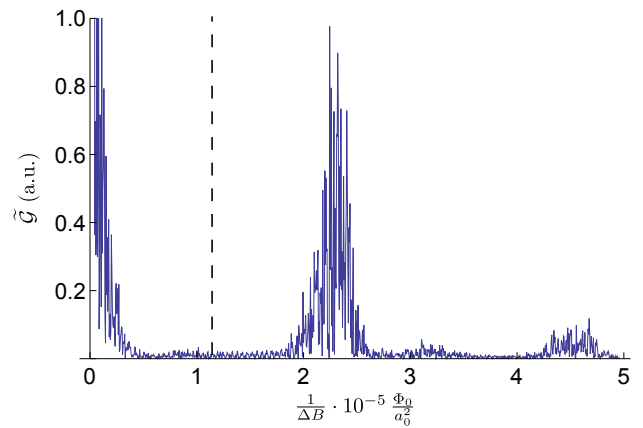}
\caption{(Color online) Frequency spectrum corresponding to Fig.~\ref{fig:magnetoconductance}, obtained from the Fourier transform $\widetilde{\mathcal{G}}$ of the magnetoconductance $\mathcal{G}$. Besides the low frequency background and the fundamental oscillation component, the second harmonic is also slightly visible in the spectrum. The dashed line indicates the frequency limit of the high pass frequency filter used for background subtraction.}
\label{fig:fourierSpectrum}
\end{figure}

In Fig.~\ref{fig:magnetoconductance}, we plot the magnetoconductance at Fermi energy $E / \tau_0 = 0.5$ and zero gate voltage ($V_g = 0$) for a particular ring realization, showing pronounced Aharonov-Bohm oscillations on top of a low frequency background. The background signal results from universal conductance fluctuations (UCF) which is typical for phase-coherent mesoscopic devices. The behavior is in agreement with the observations made in Ref.~\onlinecite{Wurm2009}, where the authors investigate an even wider magnetic field range up to the quantum Hall regime. In Fig.~\ref{fig:fourierSpectrum}, we also show the corresponding frequency spectrum obtained from a Fourier transform of the magnetoconductance. The contributions to the Aharonov-Bohm oscillations are centered around $(\Delta B \, a_0^2 \, e / h)^{-1} \sim 2.3 \cdot 10^5$. Using $\tilde{R}^2 \, \pi \cdot \Delta B = h / e$, this frequency corresponds to a mean radius $\tilde{R} / a_0 \sim 270$ of interfering electron trajectories, which perfectly lies within the boundaries of the ring.

\begin{figure}[t]
\centering
\includegraphics[width=0.95\columnwidth]{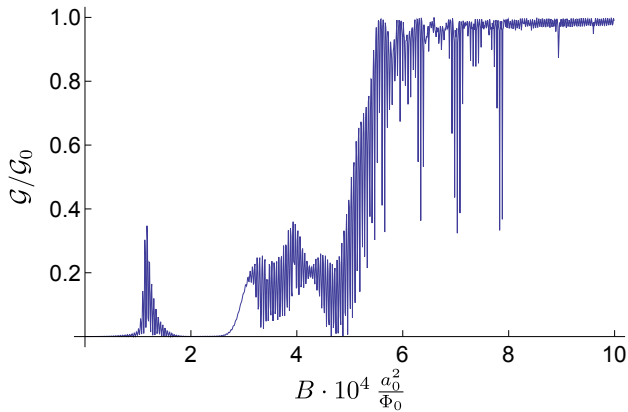}
\caption{(Color online) Magnetoconductance of a ring with $R / a_0 = 300$, $w / a_0 = 60$ at energy $E / \tau_0 = 0.1$ and zero gate voltage, showing the onset of the quantum Hall regime. Note: The conductance is still finite near zero magnetic field, which is not visible on this scale.}
\label{fig:quantumHall}
\end{figure}

In Fig.~\ref{fig:quantumHall}, we show the same plot for $E / \tau_0 = 0.1$. The oscillations diminish at $B \, a_0^2 \, e / h \sim 6 \cdot 10^{-4}$. This field strength marks the onset of the quantum Hall regime, where the cyclotron diameter becomes comparable to the width of the arms of the ring; an estimate of the graphene cyclotron diameter $d_c = 2 E / v_F e B$, taking the Fermi velocity at the Dirac point in graphene, $v_F = 3 \pi \tau_0 a_0 / h$, yields $d_c / a_0 \sim 40$, a value of same order of magnitude as the width $w / a_0 = 60$.

By applying a gate voltage $V_g > 0$ to one of the ring arms, the magnitude of the Aharonov-Bohm oscillations may be modified. A convenient measure of the oscillation magnitude is the root mean square (RMS) amplitude of the signal. Prior to the RMS analysis, the UCF background has to be removed from the signal. This is achieved by applying a high pass frequency filter to the Fourier transform of the magnetoconductance data, as indicated in Fig.~\ref{fig:fourierSpectrum}. The retained, unbiased signal is squared, and the root of the average over the squared signal is defined as the RMS amplitude $\Delta \mathcal{G}_{RMS}$.

\begin{figure}[t]
\centering
\includegraphics[width=0.95\columnwidth]{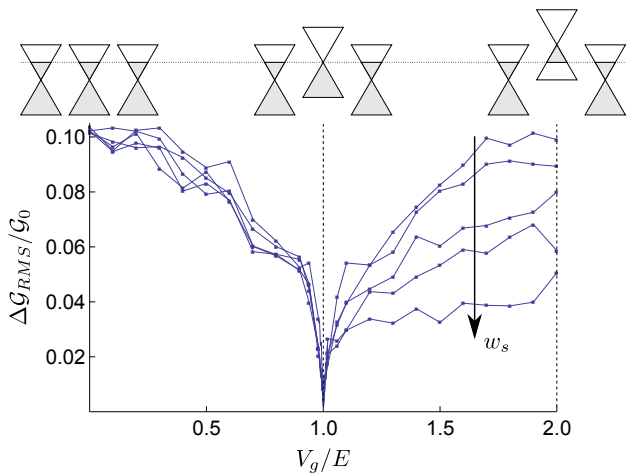}
\caption{(Color online) RMS analysis for the setup used in Fig.~\ref{fig:magnetoconductance} for different smoothing widths $w_s / a_0 \in \{ 0, 21, 52.5, 105, 210 \}$. Each data point results from an average over five realizations of edge disorder. The corresponding standard deviations lie between $0.005 \, \mathcal{G}_0$ and $0.015 \, \mathcal{G}_0$ but are suppressed for better visibility. For better clarity, the spectrum schematics (see Fig.~\ref{fig:junctionDirac}) are also included.}
\label{fig:rmsPlot}
\end{figure}

\begin{figure}[t]
\centering
\includegraphics[width=0.95\columnwidth]{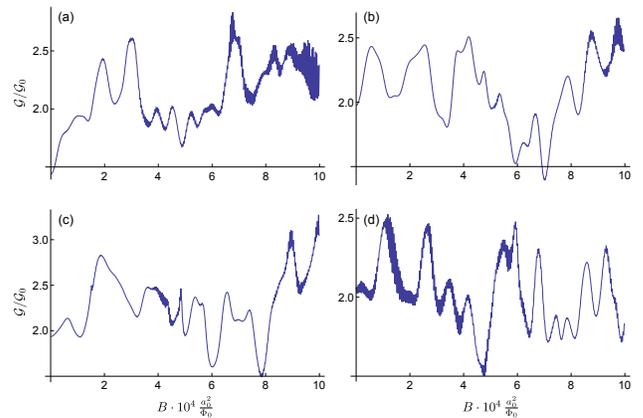}
\caption{(Color online) Magnetoconductance of a ring with $R / a_0 = 300$, $w / a_0 = 60$ tuned close to the neutrality point in the lower arm with a smooth potential interface ($w_s / a_0 = 210$). The plots (a)--(d) constitute a representative selection and are obtained by variation of the microscopic edge disorder configuration (which is always chosen randomly) for values $V_g / E = 1.00 \pm 0.01$.}
\label{fig:diracConductance}
\end{figure}

In Fig.~\ref{fig:rmsPlot} we show the dependence of the RMS oscillation amplitude $\Delta \mathcal{G}_{RMS}$ on the gate voltage $V_g$ for different smoothing widths $w_s$ (see Fig.~\ref{fig:setup}) at energy $E / \tau_0 = 0.5$, where the average is taken over the full range $B = 0 \ldots 10^{-3} \, \Phi_0 / a_0^2$. Increasing the gate voltage from zero towards the neutrality point $V_g = E$ not only leads to increased potential scattering but also to a reduction in the number of accessible propagating states in the lower arm of the ring. As can be seen in Fig.~\ref{fig:rmsPlot}, the oscillation amplitude diminishes and reaches a minimum value at the neutrality point. Note that, since the transmission through the upper ring arm is not at all affected by a gate potential, the overall conductance itself is only slightly changed to fluctuate around $2.5 \, \mathcal{G}_0$ (see Fig.~\ref{fig:diracConductance}) as compared to values around $3.4 \, \mathcal{G}_0$ in the case of zero gate potential on the lower ring arm (see Fig.~\ref{fig:magnetoconductance}).

For $V_g < E$, the decay of the RMS amplitude towards the neutrality point does not depend on the details of the gate potential interface. However, in the regime of Klein tunneling, $V_g > E$, the oscillation behavior strongly depends on the smoothness of the gate potential. In case of a smooth potential, the partial waves in the lower arm have to tunnel through a finite region of low density of states, where $V \sim E$ (see Fig.~\ref{fig:setup}), in order to interfere with the partial waves traversing the upper arm. The lower arm becomes increasingly penetrable as this region gets narrower, until it gets transparent in case of a sharp potential. This reflects the usual behavior of Klein tunneling phenomena, where the probability for tunneling through a $pn$-junction depends on the smoothness of the $pn$-interface.~\cite{Been2008,Chei2006}

The described behavior of the RMS amplitude is robust over the whole energy range under consideration, except for an increasing uncertainty at lower values for the Fermi energy. Although all results are presented for zigzag boundary conditions in the leads, the effects are independent of a change of orientation of the graphene lattice to armchair boundaries in the leads.

Before we conclude, we mention here an additional observation concerning the dependence of the magnitude of the Aharonov-Bohm oscillations on the magnetic field strength $B$, when the lower arm of the ring is tuned near the neutrality point (see Fig.~\ref{fig:diracConductance}). It seems that in this regime the oscillation magnitude is in general significantly lower for low field strength, compared to the oscillations at higher field strength. This is indeed the case for most of the ring realizations we investigated, though not for all of them (see Fig.~\ref{fig:diracConductance}(d)). The reason for such a behavior is so far not understood. Since the increase in oscillation magnitude cannot be related to any particular length scale a connection to the quantum Hall effect seems unlikely.

\section{Conclusions}
\label{sec:con}

In summary, we have numerically analyzed transport through graphene ring structures in the presence of a perpendicular magnetic field based on the recursive Green's function formalism. In order to understand the physics of the interplay of the Aharonov-Bohm effect and Klein tunneling in graphene, we have looked at the influence of a local gate over one of the arms of the ring on magnetotransport. By varying the gate voltage, we have been able to tune this arm from the $n$-type to the $p$-type transport regime via the Dirac point. The analysis of the root mean square amplitude of the Aharonov-Bohm oscillations clearly shows that the $p$-type signal is smaller for smooth $pn$-junctions in the ring arm and can recover the full $n$-type value only for very sharp $pn$-junctions. Our predictions nicely complement the analysis of Ref.~\onlinecite{Kats2010} where both arms of the ring are assumed to be tuned to the charge neutrality point. This might lead to the first observation of the Aharonov-Bohm effect caused by transport through evanescent modes or a combination of propagating modes in one arm and evanescent modes in the other arm of the ring.

\begin{acknowledgements}
We acknowledge useful discussions with C. Bruder, K. Ensslin, A.F. Morpurgo, P. Recher, C. Stampfer, M. Wimmer, and financial support from the German DFG, the Swiss NSF, and the NCCR Nanoscience.
\end{acknowledgements}

\appendix

\section{Effective description of infinite leads}\label{sec:app_effLeads}

\begin{figure}[t]
\centering
\includegraphics[width=0.95\columnwidth]{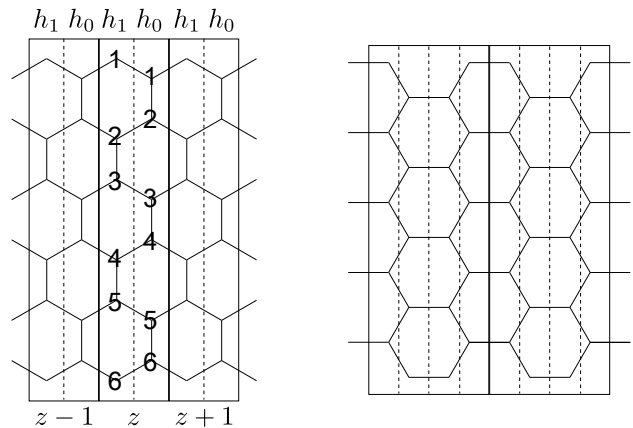}
\caption{Division of a zigzag (left) and armchair (right) lead's unit cell. In the zigzag case, the first of the two slices in which the unit cell is divided is described by $h_1$, and the second slice that contains those sites which couple to the next unit cell to the right is described by $h_0$. Analogously, in the armchair case, the first three slices of a unit cell are described by $h_1$, whereas only the fourth slice is described by $h_0$.}
\label{fig:cellDivision}
\end{figure}

The leads may be described in an effective manner if some of the atomic sites contained in cell $z$ do not couple to the adjacent cell $z + 1$. One can then conveniently divide the unit cell into two subcells according to
\begin{equation}
	H_0 \equiv \left( \begin{array}{cc}
		h_1	&	t_{10}\\
		t_{10}^\dagger	&	h_0
	\end{array} \right),
	\qquad
	H_1 \equiv \left( \begin{array}{cc}
		0	&	0\\
		t_{01} &	0
	\end{array} \right)
\end{equation}
resulting in a double-periodic structure as shown in Fig.~\ref{fig:cellDivision},
where $h_0$ only describes atomic sites that directly couple to the next cell. Writing $\psi_z = ( \tilde{\psi}'_z, \tilde{\psi}_z )^\text{T}$ then yields the effective description of the lead (\ref{for:effSG}) where
\begin{align}
	\tilde{H}_1 &\equiv t_{01} \, (E - h_1)^{-1} \, t_{10} \nonumber \\
	\tilde{H}_0 &\equiv \theta_{10} + h_0 + \theta_{01} \nonumber \\
	\theta_{10} &\equiv t_{10}^\dagger \, (E - h_1)^{-1} \, t_{10} \nonumber \\
	\theta_{01} &\equiv t_{01} \, (E - h_1)^{-1} \, t_{01}^\dagger
\end{align}
Note that simple single-periodic structures are included in the description via $h_0 = h_1, \, t_{01} = t_{10}$.

In the case of a graphene lead with zigzag edges in zero magnetic field, the non-vanishing matrix elements of the $\left( M/2 \times M/2 \right)$-matrices $t_{01}, t_{10}$ and $h_k, \, k \in \{0, 1\}$ may be written in a compact form,
\begin{align}
	(h_k)_{ii} &= V_0 \nonumber \\
	(h_k)_{ 2 i + k - 1, \, 2 i + k } &= -\tau_0 = (h_k)_{ 2 i + k, \, 2 i + k - 1 } \nonumber \\
	(t_{01})_{ii} &= -\tau_0 = (t_{10})_{ii}
\end{align}
where we assumed $\tau_0 \in \mathbb{R}$ and a constant on-site matrix element $V_0 \in \mathbb{R}$. Of course, the roles of $h_0$ and $h_1$ are interchangeable, dependent on the details of the lead's surface.

Obviously, the effective coupling matrix $\tilde{H}_1 = \tau_0^2 \, (E - h_1)^{-1}$ is invertible, in contrast to the original $\left( M \times M \right)$ coupling matrix $H_1$ in the non-effective description, which is clearly non-invertible. Additionally, the size $\tilde{M}$ of the matrices describing the effective lead is smaller by a factor of two compared to the original description, yielding increased performance, since the computational effort of the applied RGF scheme roughly scales as $\mathcal{O}(\tilde{M}^3)$. The performance gain is even larger in the case of graphene leads with armchair edges, since then only $1/4$ of the atomic sites within a particular unit cell couple to the next cell. The derivation is similar to the zigzag case and will not be repeated here.

One point to note is that $\tilde{H}_1$ is not well defined for energy values matching the eigenvalues of $h_1$, $E = V_0, \, V_0 \pm \tau_0$. $E = V_0$ corresponds to the charge neutrality point in graphene, where the density of states vanishes and hence there are no propagating modes in the lead that may contribute to the current. The singularity at $E = V_0 \pm \tau_0$ corresponds to the van-Hove singularity in graphene where the density of states diverges.

For a more formal treatment of this regularization procedure see Refs.~\onlinecite{Roch2006, Rung2008}. The subdivision of the graphene lead's unit cell as depicted in Fig.~\ref{fig:cellDivision} has also been previously used in Ref.~\onlinecite{Xu2008}.

\section{The lead--sample interface}
\label{sec:app_interface}

\begin{figure}[t]
\centering
\includegraphics[width=0.95\columnwidth]{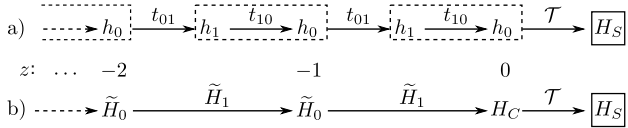}
\caption{Comparison of actual (a) and effective (b) leads, together with their coupling to a scatterer $H_S$. Effective leads couple to the scatterer through an additional contact slice, described by $H_C$.}
\label{fig:effectiveLeads}
\end{figure}

For a semi-infinite lead that terminates at $z = 0$ (see Fig.~\ref{fig:effectiveLeads}) such that the wave function on the surface of the lead is described by $\tilde{\psi}_0$ in the effective description (\ref{for:effSG}), we have to account for the boundary conditions. For a left lead, Eq.~(\ref{for:effSG}) still holds for $z < 0$ and we additionally have the equations
\begin{align}
\label{for:int1}
E \, \tilde{\psi}_0 &= t_{10}^\dagger \, \tilde{\psi}'_0 + h_0 \, \tilde{\psi}_0 + \mathcal{T} \, \psi_S\\
\label{for:int2}
E \, \psi_S &= \mathcal{T}^\dagger \, \tilde{\psi}_0 + H_S \, \psi_S
\end{align}
where $\psi_S$ and $H_S$ describe the part of the sample that couples to the lead, and $\mathcal{T}$ accounts for this coupling. By eliminating $\tilde{\psi}'_0$ as we have done before, (\ref{for:int1}) may be replaced by
\begin{equation}
E \, \tilde{\psi}_0 = \tilde{H}_1^\dagger \, \tilde{\psi}_{-2} + H_C \, \tilde{\psi}_0 + \mathcal{T} \, \psi_S
\end{equation}
where $H_C \equiv h_0 + \theta_{10}$ describes a contact slice, through which the coupling of the lead to the sample takes place in the effective description, as shown schematically in Fig.~\ref{fig:effectiveLeads}.

\begin{thebibliography}{99}
\bibitem{foot1}
Since the Green's function of the fully coupled contact slices $G_C$ yields the same transmission probabilities as $G$, one may omit this last step and calculate the conductance from $G_C$ instead.
\end{thebibliography}
\end{document}